\documentclass[%
 aip,
 amsmath,amssymb,
 reprint,%
]{revtex4-1}

\usepackage{graphicx}
\usepackage{epstopdf}

\usepackage{dcolumn}
\usepackage{bm}
\usepackage{color}
\usepackage{hyperref}
\usepackage{longtable}
\usepackage{multirow}
\usepackage{amsmath}

\begin{document}

\preprint{JAP19-AR-02647}

\title{Thermal conductivity of strained silicon: molecular dynamics insight and kinetic theory approach}

\author{Vasyl Kuryliuk}
\affiliation{Taras Shevchenko National University of Kyiv, Faculty of Physics, 64/13, Volodymrska str., 01601, Kyiv, Ukraine.}

\author{Oleksii Nepochatyi}
\affiliation{Taras Shevchenko National University of Kyiv, Faculty of Physics, 64/13, Volodymrska str., 01601, Kyiv, Ukraine.}

\author{Patrice Chantrenne}
\affiliation{Univ. Lyon - INSA Lyon - MATEIS - UMR CNRS 5510, F69621 Villeurbanne, France.}

\author{David Lacroix}
\affiliation{Universit\'{e} de Lorraine, CNRS, LEMTA, Nancy, F-54000, France.}%

\author{Mykola Isaiev}
\email{mykola.isaiev@univ-lorraine.fr}
\affiliation{Universit\'{e} de Lorraine, CNRS, LEMTA, Nancy, F-54000, France.}%
\affiliation{Taras Shevchenko National University of Kyiv, Faculty of Physics, 64/13, Volodymrska str., 01601, Kyiv, Ukraine.}%

\date{\today}

\begin{abstract}
In this work, we investigated tensile and compression forces effect on the thermal conductivity of silicon. We used equilibrium molecular dynamics approach for the evaluation of thermal conductivity considering different interatomic potentials. More specifically, we tested Stillinger-Weber, Tersoff, Environment-Dependent Interatomic Potential and Modified Embedded Atom Method potentials for the description of silicon atom motion under different strain and temperature conditions. It was shown that Tersoff potential gives a correct trend of thermal conductivity with hydrostatic strain, while other potentials fails, especially when compression strain is applied. Additionally, we extracted phonon density of states and dispersion curves from molecular dynamics simulations. These data were used for direct calculations of thermal conductivity considering the kinetic theory approach. Comparison of molecular dynamics and kinetic theory simulations results as a function of strain and temperature allowed us to investigate the different factors affecting the thermal conductivity of strained silicon.
\end{abstract}

\maketitle

\section{\label{sec:level1}Introduction}
Tuning of thermal properties of various materials is one of the key demand in material research~\cite{Volz2016, Cahill2003, Cahill2014, TermBook2017}. First and foremost, such necessity arises because of continuous miniaturization of core components of various micro-devices. As a result, issues related to the improvement of heat management are more and more crucial. In this context any possibilities of increasing or lowering the thermal conductivity in semiconductor material are very important and can encounter a broad attention in several application fields.

Strain is an effective method for tuning the thermal property of various materials due to its flexibility. Strain can affect the thermal properties of a material by shifting its phonon frequencies~\cite{Zi1992, Sui1993}, thereby changing its heat capacity, phonon group velocities, and phonon lifetimes, all of which contribute to the lattice thermal conductivity. Many attempts have been done to study the thermal conductivity of solids under strain using experiments~\cite{Andersson1988, Goncharov2012, Murphy2014} and theoretical simulations~\cite{Bhowmick2006, Xu2009, Parrish2014}. Ross et al.~\cite{Ross1984} concluded that the thermal conductivity of semiconductors increases with compressive strain. Picu et al.~\cite{Picu2003} studied strain effects on a model Lennard-Jones crystal using molecular dynamics simulations. They found that the thermal conductivity increases under compression and decreases under tension. It has been observed that the thermal conductivity decreases continuously when the strain changed from compressive to tensile for bulk silicon, silicon nanowires, and silicon thin films~\cite{Li2010}.

Besides, in addition to the control and the modification of material properties that can be expected, elastic stresses often arise as a result of  technological processing of a crystalline solid, like:  porous network formation~\cite{OMarty2006, Isaiev2015}, nanostructuration~\cite{Neimash2016, DanileFan2017}, amorphization~\cite{Newby2013, Orekhov2018}, nanoinclusion~\cite{Korotchenkov2014, Kuryliuk2015}. Additionally, other methods of heat fluxes control involving phononics membranes~\cite{Midtvedt2014}, dislocations~\cite{Ni2014, AlGhalith2016, Term2018, TaoWang2018}, functionalization of surfaces by different coatings and shells \cite{Ronggui2005, Isaiev2014} lead to generation of strongly heterogeneous fields of elastic stresses.

For these reasons a particular emphasis on the understanding of the role of strain on thermal transport properties at the nanoscale is an urgent need. As a first step in elucidating this issue, one needs more physical insight regarding phonon transport in strained materials and silicon is a good starting point to link theory and experimentation. Silicon is a material of choice in nowadays micro- and nanoelectronics. Particularly, strained silicon is one of the best candidate for implementation in metal oxide semiconductors devices~\cite{Mung2007} because of the possibility of band gap tuning with straining. Furthermore, recent studies revealed mechanical-field control of electron spin qubits of donors impurities in silicon~\cite{Mansir2018}, which open possibilities of strained silicon application in spin-based quantum technologies~\cite{Usman2018}. The change of vibration states of the silicon is also an appropriate way to tune its thermal properties.

Molecular dynamic (MD) is an important tool for predicting the thermal properties of bulk semiconductors and nanostructures, including silicon. The accuracy of the results delivered by MD simulations depends critically on the reliability of interatomic potentials. Several semi-empirical potentials have been developed for Si. The most popular Si potentials were proposed by Stillinger and Weber (SW)~\cite{Stillinger1985} and Tersoff~\cite{Tersoff1988}. Other Si potential formats include the Environment-Dependent Interatomic Potential (EDIP)~\cite{Justo1998}, and the Modified Embedded Atom Method (MEAM)~\cite{Baskes1989} potentials. These original potentials were modified by many authors by slightly changing the analytical functions and improving the parametrization~\cite{Tersoff_1988, Tersoff1989, Tersoff1990, Lee2000, Lee2001}. Several studies  have  been  conducted  to  compare different empirical interatomic potentials for heat transport modeling  in silicon~\cite{AbsdaCruz2011, Howell2012, El-Genk2018}; however, no study has systematically compared potentials regarding strain effect on the thermal conductivity of bulk silicon.  Therefore, the first goal of this work is to test the Tersoff~III~\cite{Tersoff_1988, Tersoff1989, Tersoff1990}, SW~\cite{Stillinger1985}, EDIP~\cite{Justo1998}, and Second Nearest Neighbour (2NN) MEAM~\cite{Lee2000, Lee2001} potentials for their ability to predict the the thermal conductivity of silicon under hydrostatic compressive and tensile strains. Interatomic parameters of each considered potentials are provided in the Appendix~\ref{Potentials}.

In this study, we used equilibrium molecular dynamics (EMD) approach for simulation of thermal conductivity of strained silicon. Since thermal conductivity evaluated with molecular dynamics is sensitive to the interatomic interactions~\cite{AbsdaCruz2011, Howell2012, El-Genk2018}, we chose the optimum potential for simulation of thermal conductivity of strained Si silicon. With the chosen potential, we calculated dispersion curves, density of states and phonon lifetimes for different temperatures and strains. However, the significant drawback of MD simulations is the classical description, which induce no energy quantization. Therefore, we applied additionally analytic approach based on the kinetic theory (KT) of gases~\citep{Chantrenne2005} for evaluation of the thermal conductivity with the input data calculated by molecular dynamics. Finally, we investigated correlations between the temperature dependence of thermal conductivity revealed by KT simulations with the data obtained by EMD.

\section{\label{sec:level1} Simulation Methods}
\subsubsection{Formation of strained silicon and elastic properties calculation}
All MD simulations were carried out with Large-scale Atomic/Molecular Massively Parallel Simulator (LAMMPS). The initial system was set as a single crystal silicon with a diamond lattice structure. The initial lattice parameter was set to be equal to 5.43~\AA. The $x$, $y$ and $z$ dimensions of the silicon slab as well as the simulation domain was set to be equal to 10 lattice cells. It has been shown that with these cell sizes, the system-size effects can be eliminated~\cite{Sellan2010, Wang2017}. Additionally, we tested the impact of the simulations domain size on the strain-stress curves, and found it insignificant (see Appendix~\ref{strain-stress-dif-cells}). The periodic boundary conditions were used in all directions to simulate a bulk crystal. The system was equilibrated under temperature $T$ in the Nos\'{e}-Hoover thermostat with the use of isotherm-isobaric (NPT) ensemble during 100~ps with time step equal to 1 fs to achieve equilibrium lattice parameter for each temperature and potential for the case of external pressure absence ($P$=0~Pa). The initial systems after equilibrium were used for thermal conductivity calculations of pristine Si structures.

Then, the pristine systems were slowly hydrostatically deformed in compression or tension. The rate of strain was set to be equal to $10^{-3}$~ps$^{-1}$. After each 0.1~ps of deformation procedure the information regarding strain and induced stress in the system was collected. Such deformation was performed during 200~ps, and the final deformation was equal to $\pm 20 \%$ to achieve stress-strain curve visualization. Additionally, for the strains equal to $\pm 1 \%, \pm 2 \%, \pm 3 \%, \pm 4 \%$, and $\pm 5 \%$ the re-start files with atoms positions and velocities were collected. These files were used for further thermal conductivity evaluation.

\begin{figure}[h]
	\includegraphics[width=\columnwidth]{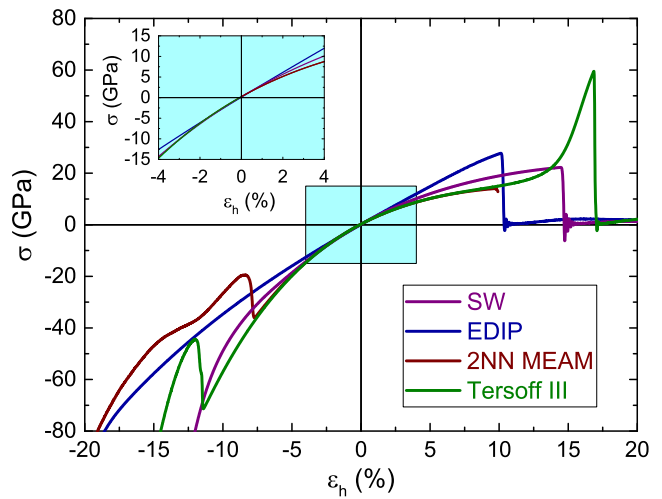}
	\caption{\label{strain_stress} Strain-stress curves calculated for different potentials, EDIP, Tersoff~III, and 2NN MEAM at 300~K, negative strain stands for compression and positive for tension.}
\end{figure}

Fig.~\ref{strain_stress} depicts the typical strain-stress curves of a silicon sample in compression/tensile tests for the four considered potentials at 300~K. As one can see from the figure, bulk modulus ($B$) can be estimated from the slope of a curve divided on 3 in the elastic deformation range (blue window in Fig~\ref{strain_stress}). For all potentials $B$ values are the same and are equal to $B=100\pm8$~GPa. The latter value matches well experimental data~\cite{Hopcroft2010,MatProp}. High frequency oscillations on strain-stress curves at large tensile deformations are related to material fracture. The abrupt changes of strain-stress curves obtained with Tersoff~III and 2NN MEAM potentials at high compressive strain are caused by phase transformation in silicon under hydrostatic pressure~\cite{Mizushima1994, Yang2004}.

\subsubsection{Thermal conductivity evaluation}
For the thermal conductivity evaluation, we used the Equilibrium Molecular Dynamics (EMD) approach, which is based on Green-Kubo formalism. In this framework, thermal conductivity can be evaluated from auto-correlation of heat flux as follows~\cite{McGaughey2004}:
\begin{equation}\label{Green_Kubo}
	\kappa_{ij} = \frac{1}{V k_B T^2}\int_{0}^{\infty}{dt\left<J_i(t)J_j(0)\right>}
\end{equation}
where $V$ is the volume of the system, $k_{B}$ is Boltzmann's constant, and $\left<J_i(t)J_j(0)\right>$ is the heat
current autocorrelation function. In general, the conductivity is averaged across all three directions to obtain the scalar
conductivity.

Since EMD is based on numerical simulations, the integration time as well as the averaging time in Eq.~\ref{Green_Kubo} were limited to finite values. Additionally, the thermal conductivity was averaged in different directions.  Thus, Eq.~\ref{Green_Kubo} was used in the following form:

\begin{equation}\label{Green_Kubo_mod}
	\kappa = \frac{1}{3 V k_B T^2}\int_{0}^{t_{c}}{dt\left<\textbf{J}(t)\textbf{J}(0)\right>_{t_{s}}}
\end{equation}
where $\textbf{J}$ is the heat flux vector, $t_{c}$ is a finite correlation time for which the integration is carried out, and $t_{s}$ is the sampling time over which the autocorrelation function is accumulated for averaging.

The atom structure input files of EMD calculation of pristine and strained samples are the ones described above. For thermal conductivity calculation, the systems were equilibrated again according to  Nos\'{e}-Hoover thermostat during 1~ns, but with the use of canonical (NVT) ensemble. After equilibrium, we performed NVE integration during 25~ns. During this integration, the heat fluxes in different directions were collected and recorded. Thermal conductivities were extracted with the Eq.~\ref{Green_Kubo_mod}. The correlation time was chosen to be equal 10~ps, and the sampling time was 50~ps. It should be noted that we tested also sampling time equal to 100~ps to check stability of calculations. In this case, duration of NVE integration was equal to 100~ns. Particularly, thermal conductivity for Tersoff potential under 300~K with longer sampling time (100~ps) was found to be equal $\kappa$ = $229\pm8$~W/(m K), and for shorter sampling time (50~ps) $\kappa$ = $222\pm8$~W/(m K)). Thus, both values are in good agreement, and therefore for further calculation we used only the shorter sampling time equal to 50~ps. Since, thermal conductivity calculated based on Green-Kubo formalism is sensitive to initial conditions, the simulations were carried for 10 different seeds. The final value of thermal conductivity for each potential, strain and temperature was averaged among these seeds.

\subsubsection{Calculation of phonon dispersion and density of states}
To get more details about strain effect on the silicon thermal conductivity, the phonon density of states (DOS) and the phonon dispersions of silicon were calculated. Those calculations were performed using the FixPhonon module of LAMMPS. With this tool, dispersion curves can be obtained from the Green's function formalism \cite{Kong2009, Kong2011}. In the latter, we consider the lattice vibrations at finite temperature $T$; the $k^{\rm th}$ basis atoms in the $l^{\rm th}$ unit cell are displaced from their equilibrium positions $\textbf{r}_{lk}$ by an amount of $\textbf{u}_{lk}$ (atom displacement). In the reciprocal space, the displacements are obtained from the Fourier transformation of the real space ones ~\cite{Kong2011} according to :
\begin{equation}\label{displ}
	\mathbf{\tilde{u}}_{k\alpha}(\mathbf{q})=\dfrac{1}{\sqrt{N_{UC}}}\sum_{l}\mathbf{u}_{lk\alpha}e^{-i\mathbf{q}\mathbf{r}_{l}}
\end{equation}
where $N_{UC}$ is the total number of unit cells in the crystal, $\mathbf{r}_{l}$ is the equilibrium positions of $l^{\rm th}$ unit cell, $\alpha$ is the component of the atomic displacement of the $k^{\rm th}$ atom in Cartesian coordinates, and $\mathbf{q}$ is the wave vector. The Green's function in reciprocal space given by :
\begin{equation}\label{GreenF}
	\mathbf{\tilde{G}}_{k\alpha,k'\beta}(\mathbf{q})=\langle \mathbf{\tilde{u}}_{k\alpha}(\mathbf{q})\mathbf{\tilde{u}^{\ast}}_{k'\beta}(\mathbf{q}) \rangle
\end{equation}
where superscript $\ast$ denotes the complex conjugate and $\langle \dots \rangle$ denotes the ensemble average. With the Green's function method, the dynamical matrix can be expressed as:
\begin{equation}\label{DynMatr}
	\mathbf{D}_{k\alpha,k'\beta}(\mathbf{q})=\dfrac{k_{B}T}{M}[\mathbf{\tilde{G}}^{-1}(\mathbf{q})]_{k\alpha,k'\beta}
\end{equation}
Here $M$ is the mass of Si atom. By solving the eigenvalues of dynamical matrix $\mathbf{D}$, we can get the frequencies of all phonon modes:

\begin{equation}\label{EigenVal}
	|\mathbf{D}_{k\alpha,k'\beta}(\mathbf{q})-\delta_{\alpha\beta}\delta_{kk'}\omega^{2}(\mathbf{q})|=0
\end{equation}
where $\omega$ is the phonon frequency. Hence, the relations between $\mathbf{q}$ and $\omega$ are obtained. The phonon DOS curve is computed from the phonon dispersion by dividing the frequency range into many small segments and counting the number of states in each segment.

In this work, the dynamic matrix calculations were performed in the NVE ensemble with a Langevin thermostat. The systems were equilibrated for $0.5\cdot10^{6}$~time steps, and then system is run in the NVE ensemble for $6\cdot10^{6}$ time steps to record the atomic displacements and velocities. Using the eigenvalues of the dynamical matrices, the phonon DOS and the phonon dispersions were then calculated using the auxiliary post-processing code $phana$~\cite{Kong2011}.

\subsubsection{Calculation of phonon relaxation times}
The vibrational lifetime was estimated from the recently developed method based on the Monte Carlo-based moments approximation~\cite{Dickel2010-1, Dickel2010-2, Gao2014}. Beginning with the harmonic force constant matrix, the normal modes are found, indexed by wave vector $q$ and branch $b$. The calculation involves ensemble averaging of products of normal mode amplitudes $A_{b}$ and accelerations $\ddot{A_{b}}$, which are obtained by projecting the atomic displacements and forces onto the normal modes. The mode lifetimes are obtained from~\cite{Gao2015}
\begin{equation}\label{tau_phonon}
	\tau_{q}=\frac{1.41}{\sqrt{\frac{\langle A_{q}\ddot{A_{q}}\rangle}{\langle A_{q}^{2}\rangle}}\left(\frac{\langle \ddot{A}_{q}^{2}\rangle \langle A_{q}^{2}\rangle}{\langle A_{q}\ddot{A_{q}}\rangle^{2}}-1\right)}
\end{equation}
Calculations of phonon relaxation times were carried out with $Jazz$ python wrapper for LAMMPS~\cite{Gao2015,Jazz}, implemented to calculate the lifetimes of vibrational normal modes.

\section{\label{sec:level2} Elementary kinetic theory approach for phonons propagation}
\subsubsection{Thermal conductivity evaluation}
In this section, we will describe elements of kinetic theory approach for phonons propagation to calculate the thermal conductivity of strained Si. For the calculation, we use the analytical model developed previously by P. Chantrenne et al~\cite{Chantrenne2005}.  The model considers phonons to be particles that follow Bose-Einstein statistics. The thermal conductivity $\kappa$ in the direction $z$ associated with the phonons $(q, p)$ can be written as
\begin{equation}
	\kappa_{z}(q,p) = C(q,p) v^{2}(q,p)\tau(q,p)\cos^2[\theta_z(q)] \\
\end{equation}
where $q$ is the wave vector, $p$ its polarization, and $v$ the
group velocity determined from the dispersion curves:
\begin{equation} \label{group_velocity}
	v = \frac{d\omega(q,p)}{dq}
\end{equation}
with $v$ the angular velocity, $\tau(q,p)$ is the phonon relaxation
time due to the phonon-scattering phenomena, $\theta_z(q)$ is the
angle between the wave vector $q$ and the direction $z$, and
$C(q,p)$ is the specific heat per unit volume.
\begin{equation}
	C(q,p) = k_B x^{2} \frac{e^{x}}{V (e^{x}-1)^{2}}
\end{equation}
\begin{equation}
	x= \frac{\hbar \omega(q,p)}{k_B T}
\end{equation}
The total thermal conductivity is the sum
of the individual contributions due to all the wave vectors $q$
and polarizations $p$:
\begin{equation}\label{LambdaSum}
	\kappa_z = \sum\limits_q \sum\limits_p \kappa_z(q,p)
\end{equation}
In our case, the system is isotropic and the size of it can be considered infinite. Thus, the sum in Eq.~\ref{LambdaSum} becomes an integral. It is expressed in terms of angular frequency:
\begin{multline} \label{KinTheorChartrenne}
	\kappa =
	6 \frac{k_B}{3 \frac{a^3}{4}} \int\limits_0^{\infty} d\omega (\frac{\hbar\omega}{k_B T})^2 \frac{e^\frac{\hbar\omega}{k_B T}}{(e^\frac{\hbar\omega}{k_B T} - 1 )^2} D(\omega) \times \\
	\times\sum\limits_p \xi(\omega, p) v^2(\omega, p) \tau(\omega,p)
\end{multline}
where 6 is the number of polarizations, $\frac{a^3}{4}$ is the volume of a primitive cell, $D(\omega)$ is the phonon density of states ($\int\limits_0^\infty D(\omega)d\omega = 1$), $\tau(\omega,p)$ is the phonon relaxation time, $\xi(\omega,p)$ are coefficients, that express the contributions of different polarizations to DOS.
\begin{equation}
	\begin{split}
		\xi(\omega,p) \propto \frac{q^2 (\omega,p)}{v(\omega,p)} \\
		\sum\limits_p \xi(\omega,p) = 1
	\end{split}
\end{equation}

\subsubsection{Relaxation times}
In order to theoretically assess thermal conductivity, the last needed parameter is the phonon relaxation time. In general, several scattering mechanism are at play. According to Mathiessen rule, the resulting relaxation time (RT) is
\begin{equation}
	\tau^{-1}(\omega, p) = \tau^{-1}_{U}(\omega, p) + \tau^{-1}_{BC}(\omega, p) + \tau^{-1}_{d}(\omega, p)
\end{equation}
where  $\tau^{-1}_{U}(\omega, p)$ is RT due to Umklapp scattering, $\tau^{-1}_{BC}(\omega, p)$ is RT due to the presence of system boundaries and $\tau^{-1}_{d}(\omega, p)$ is RT due to defect or impurity scattering. In our case, the system is considered as infinite (periodic boundary conditions set in EMD), so $\tau^{-1}_{BC}(\omega, p)=0$ , and the lattice is supposed to be perfect, so it has no defects. Thus,
\begin{multline}
	\tau^{-1}(\omega, p) = \tau^{-1}_{U}(\omega, p) = \\ =
	\begin{cases}
		A_T \omega^{\chi_T} T^{\xi_T} \exp(\frac{B_T}{T})\quad  \text{TA branches} \\
		A_L \omega^{\chi_L} T^{\xi_L} \exp(\frac{B_L}{T})\quad \text{LA branches}
	\end{cases}
\end{multline}
Besides, it should be mentioned that the implemented approach for calculations of relaxation times only gives a global  relaxation time. Decomposition of different contribution is a challenging issue and is not the purpose of this work. Therefore, we fitted numerical data evaluated with Eq.\ref{tau_phonon} by the following equation for all polarizations:
\begin{equation} \label{fittau}
	\tau(\omega, p) = A^{-1} \omega^{-\chi} T^{-\xi} \exp({-B/T})
\end{equation}
The fitting was performed to overcome issues arised because of the significant scattering of data evaluated by MD (see Fig.~\ref{Phonon_lifetimes}). The fitting parameters are presented in Appendix~\ref{fit-phl} (Table~\ref{Fit_Table}).

\section{\label{sec:level2} Results and Discussion}
\begin{figure}[h]
	\includegraphics[width=\columnwidth]{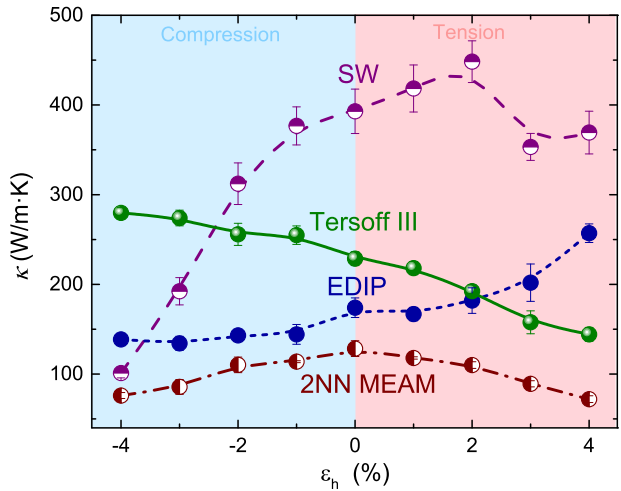}
	\caption{\label{k_vs_Strain} Thermal conductivity of silicon as a function of hydrostatic strain for SW potential, EDIP, Tersoff~III, and 2NN~MEAM at 300~K.}
\end{figure}

In this section, the results of calculations of the thermal conductivity $\kappa$ of Si under hydrostatic strain $\varepsilon_{h}$ using SW, EDIP, 2NN~MEAM and Tersoff~III potentials at $T=300$K will be discussed. The strain-dependencies of the thermal conductivity predicted with the use of these potentials are plotted in Fig.~\ref{k_vs_Strain}. Firstly, it should be noted that thermal conductivity evaluated with MD simulations overestimates the experimental results for natural~\cite{Glassbrenner1964} and highly isotopically enriched~\cite{Inyushkin2018} Si ($\kappa$=130-160~W/(m$\cdot$K) for 300 K). As discussed previously, MD is based on classical laws of motion and quantum phenomena are not taken into account. This behavior in well known, however in further analysis we will  only focus on the dependence of the thermal conductivity versus applied strain as a main parameter for the chosen potentials reliability.

Fig.~\ref{k_vs_Strain} shows that calculated thermal conductivity of unstrained SW silicon is about 393~W/m$\cdot$K. Under compression strain, the thermal conductivity of SW silicon rapidly decreases, which contradicts previously published  MD simulation results for bulk silicon~\cite{Li2010} and bulk argon~\cite{Picu2003}. For a tensile strain, the SW potential leads to non-monotonic dependence of thermal conductivity, which is not in agreement with other data already revealed by MD method~\cite{Li2010,Picu2003}. Therefore, the use of such potential is clearly inappropriate for the simulation of the thermal conductivity of silicon under hydrostatic strain.

Next, the value of $\kappa$ resulting from the EDIP calculations for unstrained silicon was found to be about 174~W/(m$\cdot$K). This potential gives also non-physical trends $\kappa(\varepsilon_{h})$: the thermal conductivity increases continuously from compression strain to tensile hydrostatic strain, which is in a sharp contrast to the expected dependence.

Despite, 2NN~MEAM potential (see Fig.~\ref{k_vs_Strain}) gives a correct strain dependence of $\kappa$ with the increase of tensile strain, it gives an insignificant increases with the increase of compression strain. Therefore, the 2NN~MEAM potentials may not be suitable for studying the strain effect on the thermal conductivity of silicon.

Finally, the calculated value of $\kappa$ for unstrained silicon with Tersoff~III~potential is about 229~W/(m$\cdot$K). This is in a good agreement with other MD results~\cite{Lee2007}. It can be noticed that the thermal conductivity of Tersoff silicon decreases continuously when the applied strain is changed from compressive to tensile. For a 4$\%$ tensile hydrostatic strain, there is a 37$\%$ reduction (from 229~W/(m$\cdot$K) to 144~W/(m$\cdot$K)) in thermal conductivity. On the other side, at a 4$\%$ compressive hydrostatic strain, the thermal conductivity increases 22$\%$ to 280~W/(m$\cdot$K). This result qualitatively corresponds to the experimental data obtained by Andersson and Backstrom~\cite{Andersson1988} for the thermal conductivity of silicon under hydrostatic pressure. According to their experiments, the thermal conductivity $\kappa$($P$) slowly increases at room temperature with hydrostatic pressure $P$ as $\kappa$/$\kappa_{0}$=1+0.004$P$, where $\kappa_{0}$ is the thermal conductivity of unstrained silicon and $P$ being measured in GPa. From Fig.~\ref{strain_stress}, the strain of -4$\%$ corresponds to the hydrostatic pressure about of 15 GPa. Therefore, the expected increasing of the thermal conductivity according to this equation would be equal to 6$\%$, which is smaller than one calculated with the Tersoff potential. More importantly we notice that the general trend of $\kappa(\varepsilon_{h})$ variations obtained with Tersoff~III potential, plotted in Fig.~\ref{k_vs_Strain}, is similar to the previously reported strain effects on the thermal conductivity of bulk silicon~\cite{Andersson1988, Li2010, Xu2009} and Si nanostructures~\cite{Yang2014, Shahraki2015}.

The results shown in Fig.~\ref{k_vs_Strain} indicate that only Tersoff~III potential is reliable for thermal conductivity calculations of Si under hydrostatic strain, while SW, EDIP and 2NN MEAM empirical potentials fails in comparison with experimental~\cite{Andersson1988} and DFT results~\cite{Parrish2014}, especially for compressive strain. Such mismatching arises because these potentials were parametrized to reproduce physical properties of some stable structures of silicon. Thus, these potentials could not describe correctly the phase transformation considering the latter specific parametrization. Silicon undergoes a phase change from the diamond lattice structure to the $\beta$-Sn lattice structure at $T$=300 K when compressive stresses becomes greater than 12 GPa~\cite{Yang2004}. This transition, using MD and Tersoff potential, is observed for larger pressure (60 GPa)~\cite{Mizushima1994} than the one experimentally measured.

The first possible explanation for the differences between the potentials in a predicting the $\kappa$($\varepsilon_{h}$) dependencies of silicon under hydrostatic compression is the wide range of cutoff radius $r_{c}$ of each potential. The corresponding value $r_{c}$ is 3.77{\AA} for SW potential, 3.12{\AA} for EDIP potential, 4.3{\AA} for 2NN MEAM potential, and 3.0{\AA} for Tersoff~III potential. Since these empirical potentials are optimized for a particular local environment of atomic configurations, they are generally not designed for conditions where the number of interacting neighbors can change abruptly. In fact, when the atomic structure is strongly compressed, the the number of neighboring atoms taken into account in the energy calculation may drastically change, leading to sharp energy variation. The fact that the Tersoff~III potential is applicable over a relatively wide range of compressive strain is due to its short cutoff range.

Additionally, different functional forms and the number of fitted parameters also lead to various thermal conductivity behavior when modifying strain. For example, the SW potential is a linear combination of two- and three-body interaction terms that stabilize the diamond cubic structure of crystalline silicon. It has only eight independent parameters and it is fitted to reproduce few experimental properties of both crystallized and liquid silicon. The Tersoff~III functional form includes many-body interactions thanks to a bond order term. This potential has 13 adjustable parameters, which are determined by heat of formation, cohesive energy, bulk modulus and the relative stability of various Si polytypes obtained from calculations and experiments. The EDIP potential has a functional form similar to that of Tersoff but slightly more complicated. Thirteen parameters are determined by fitting to a fairly small ab-initio database. MEAM is an angle dependent functional form that evolved out of the simpler embedded atom method. Because parameter values of Tersoff~III potential were determined by fitting the elastic constants~\cite{Tersoff_1988}, the potential is more likely to be applicable for the analysis of deformation behavior. The latter suggests that the Tersoff~III potential would be a reliable choice for describing Si thermal transport properties under hydrostatic strain. So further consideration of this work will be carried out with this empirical potential.

\begin{figure}[h]
	\includegraphics[width=\columnwidth]{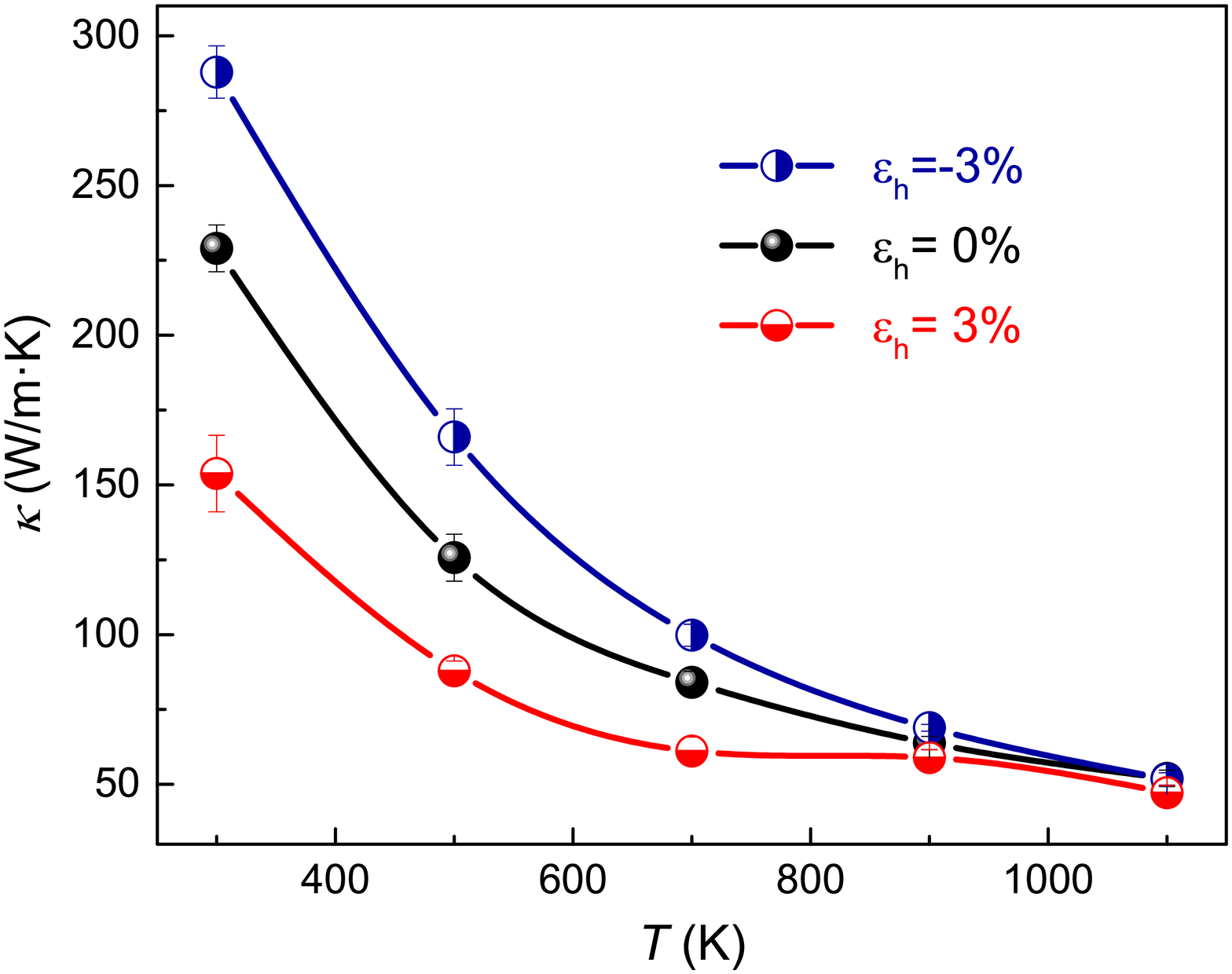}
	\caption{\label{k_vs_Temperature} Strain-dependent bulk thermal conductivity of silicon between 300 and 1100~K for Tersoff~III potential.}
\end{figure}

Figure~\ref{k_vs_Temperature} shows the calculated thermal conductivity of bulk Si crystal with the use of Tersoff~III potential under hydrostatic compressive and tensile strains at the temperature range 300-1100 K comparing to the results of the unstrained case. The results indicate that, for hydrostatic compression, the strain field increases the thermal conductivity and enables to affect the conductivity in the whole range of temperature from 300 to about 900~K, while the tensile hydrostatic strain decreases the thermal conductivity of silicon. For the temperatures over $\sim$900~K, the strain effects on the thermal conductivity becomes weak enough to be negligible.

Eventually, we present further analysis of the thermal conductivity variation with hydrostatic strain and temperature based on the Eq.~\ref{KinTheorChartrenne}. From this equation, we  can consider the changes in phonon density of states, phonon group velocity, and phonon relaxation time to understand the effect of hydrostatic strain on the thermal conductivity of Si. Moreover Eq.~\ref{KinTheorChartrenne} takes into account the energy distribution on the phonon mode as a function of the temperature level.

\begin{figure}[h]
	\includegraphics[width=\columnwidth]{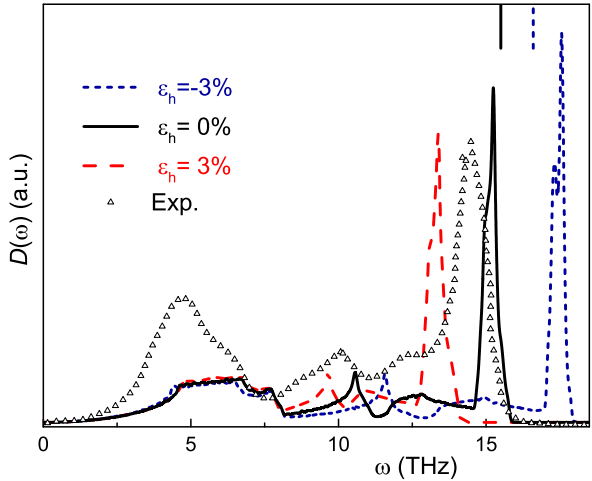}
	\caption{\label{DOS_vs_Strain} Phonon DOS of silicon calculated with Tersoff~III potential at 300 K under different hydrostatic strains: -3$\%$, 0$\%$, and 3$\%$. Triangles corresponds to the experimental dependence $D$($\omega$) obtained by inelastic neutron scattering~\cite{Kim2015}. The vertical lines at the top represents the experimental peak positions of LO phonon mode in unstrained (solid line) and compressed at 3$\%$ (dashed line) bulk Si adopted from Reference [~\cite{Weinstein1975}].}
\end{figure}

First, Fig.~\ref{DOS_vs_Strain} shows the strain effect on the phonon DOS $D$($\omega$) for bulk and strained silicon at 3$\%$ compressive and tensile strains. The calculations with the Tersoff~III potential for unstrained silicon satisfactorily reproduce both the structure of the experimental dependence $D$($\omega$) and the frequency positions of phonon modes. It is shown that the optical phonon modes show a red shift when the tensile strain is applied, while a blue shift is observed for the compressive strain. At a 3$\%$ tensile hydrostatic strain, the frequency of the optical phonon mode changes from 15.3 to 13.4~THz ($\Delta \omega$=-1.9~THz), while at a 3$\%$ compressive hydrostatic strain, the frequency shift is $\Delta \omega$=2.2~THz. The last result correlates with the experimental Raman spectra obtained by Weinstein and Piermarini~\cite{Weinstein1975} for silicon under the hydrostatic pressure. According to their experiments, the frequency $\omega_{LO}$ of the LO phonon mode in Si shifts with pressure $P$ as $\omega_{LO}$=15.57+0.16$P$-0.002$P^{2}$ [THz]. The spectral positions of the LO phonon bands obtained by using this relation for pristine and compressed at 3$\%$ silicon are shown by the straight vertical lines in Fig.~\ref{DOS_vs_Strain}. The expected frequency shift from the experimental data is about 1.1 THz and is smaller than one obtained from MD simulations with Tersoff~III potential.

Consequently, optical phonons will have a lower or higher energy in comparison to the unstrained case when strain is applied. There are similar peak shifts for longitudinal acoustic (LA) phonons, indicating that the hydrostatic strain has a significant effect on them as well. In particular, the frequency shift for LA phonons peak in DOS is $\Delta \omega$=-0.9~THz (from 10.6~THz to 9.7~THz) and $\Delta \omega$=1.0~THz for 3$\%$ tensile and compressive hydrostatic strains, respectively. On the other hand, variations in transverse acoustic (TA) phonons are not significant compared to the LA and optical phonons. It is known that optical phonons contribute little to heat transfer~\cite{Esfarjani2011, Larkin2013}, so strain-induced transformations of phonon DOS may affect the thermal conductivity of silicon mostly due to frequency shift of the LA phonon modes.

In addition, our calculations show that temperature has little effect on the phonon DOS (not presented in Fig.~\ref{DOS_vs_Strain}). For example, the frequency shift for LA phonons peak in DOS is only $\Delta \omega$=-0.3~THz, when temperature changes from 300 to 1100~K.

\begin{figure}[h]
	\includegraphics[width=\columnwidth]{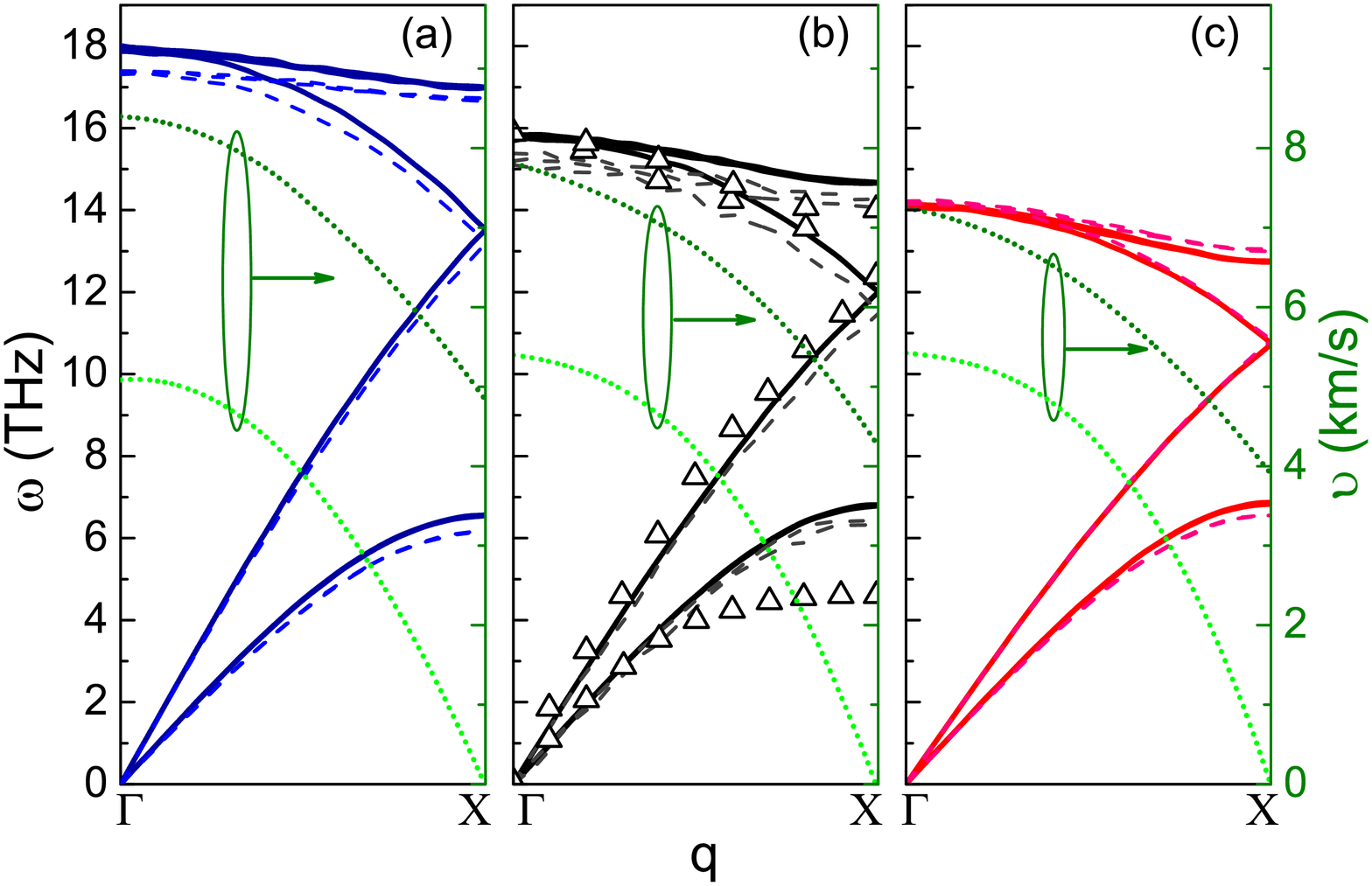}
	\caption{\label{Disp_vs_Strain} Phonon dispersion curves of silicon calculated with Tersoff~III potential at 300~K (solid lines) and 1100~K (dashed lines) under different hydrostatic strains: (a) -3$\%$ , (b) 0$\%$ , and (c) 3$\%$. Triangles corresponds to the experimental data from Reference [~\cite{Flensburg1999}]. The dotted lines represent the phonon group velocities, obtained from the dispersion curves for longitudinal and transverse acoustic phonons.}
\end{figure}

The changes in phonon group velocity can be obtained from the dispersion curves, which are plotted in Fig.~\ref{Disp_vs_Strain} for the typical hydrostatic strains ( -3$\%$, 0$\%$, and 3$\%$). The dispersion curves for unstrained Si are compared to experimental results, adopted from work[~\cite{Flensburg1999}]. Tersoff~III overestimate TA modes by more than 40$\%$ for all large wave vectors, which leads to a logical increase in thermal conductivity compared to experimental one. As shown in Fig.~\ref{Disp_vs_Strain}, as the strain condition changes from compressive to tensile, transverse acoustic phonon branches change little, while the frequency of optical and longitudinal acoustic phonon branches decrease and whole range of frequency becomes narrowed. This agrees with the change of DOS presented in Fig~\ref{DOS_vs_Strain}. Decreasing of the maximum phonon frequency as the strain moves from compressive to tensile leads to decreasing the phonon group velocities.

The evaluated phonon group velocities are plotted in Fig.~\ref{Disp_vs_Strain} by dotted line curves. As one can see, strain does not affect the TA phonon group velocity, while the LA one decrease as the applied strain is changes from compressive to tensile. These observations are consistent with our EMD thermal conductivity calculations presented above (Fig.~\ref{k_vs_Temperature}). When the Si structure is under compression, LA phonon frequencies are larger as well as group velocities. It results in a larger phonon heat flux and thus in a bigger thermal conductivity as compared to pristine case. Oppositely, traction of Si structure reduces maximum frequencies and related group velocities and thus heat flux and thermal conductivity. In addition, our calculations show that the temperature has a negligible effect on the dispersion curves of silicon and, consequently, on the phonon velocities. This is confirmed, in Fig.~\ref{Disp_vs_Strain} the dashed lines show the calculated dispersion curves for silicon at $T$=1100~K. For example, our calculations show (not presented in Fig.~\ref{Disp_vs_Strain}), that increase in temperature from 300 to 1100 K leads to changes in group velocity only $\Delta \upsilon\leq$2-4$\%$ for transverse acoustic phonons, and $\Delta \upsilon\leq$1$\%$ for longitudinal acoustic phonons in a wide range of wave vectors.

\begin{figure}[h]
	\includegraphics[width=\columnwidth]{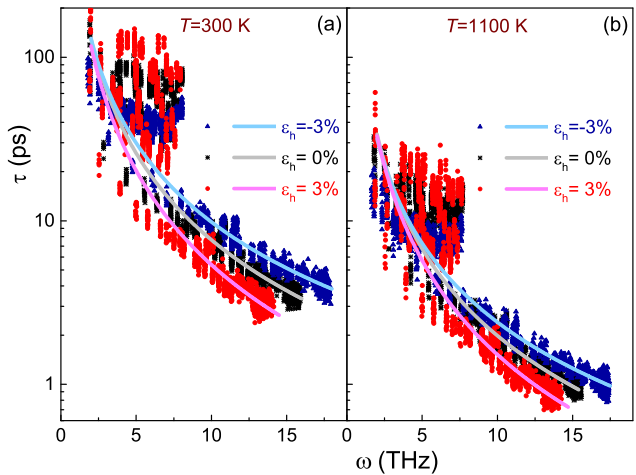}
	\caption{\label{Phonon_lifetimes} Phonon lifetimes of silicon calculated with Tersoff~III potential at 300 K (a) and 1100 K (b) under different hydrostatic strains: -3\%, 0\%, and 3\%. The solid lines correspond to fitting of the simulated data with the Eq.~\ref{fittau}.}
\end{figure}

In Fig.~\ref{Phonon_lifetimes}, we present the frequency dependent phonon lifetimes in pristine and strained silicon at 300~K and 1100~K. As one can see, lifetimes $\tau$ varies insignificantly as the system moves from compressive to tensile hydrostatic strain. For example, the mean phonon lifetimes for LA phonons with frequency $\omega\simeq$10 THz are about $\tau$=7 ps, $\tau$=8 ps, and $\tau$=6 ps in pristine, compressed at -3$\%$, and tensed at 3$\%$ silicon, respectively. The strain induced changes in phonon lifetimes for TA modes are also insignificant, so reduction of phonon mean free path as the system moves from compressive to tensile hydrostatic strain is mostly associated with the decreasing of the group velocities described above. Additionally, Fig.~\ref{Phonon_lifetimes} shows a significant (more than three times) reduction of the phonon lifetimes over almost whole frequency range when temperature increases from 300~K to 1000~K. This reduction results from increasing phonon scattering rate due to phonon-phonon interaction.  Therefore, the trend shown on Fig~\ref{k_vs_Temperature} for the temperature dependence of the thermal conductivity is mainly associated with changes in phonon lifetimes. Finally, this analysis shows that the strain and temperature dependence of the thermal conductivity of silicon arises from the change of the phonon DOS, phonon group velocity, which monotonically decreases when strain goes from compressive to tensile, and the temperature dependence of the phonon relaxation time.

\begin{figure}[h]
	\includegraphics[width=\columnwidth]{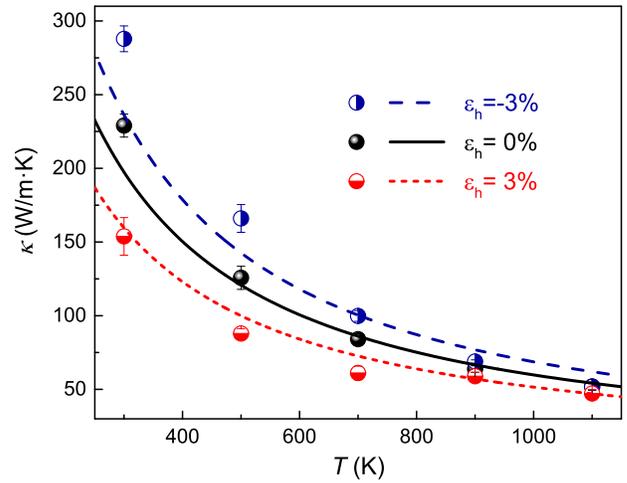}
	\caption{\label{k_vs_Temperature+Analytic} Strain-dependent bulk thermal conductivity of silicon between 300 and 1100~K; circles: EMD simulation results; lines: kinetic theory results.}
\end{figure}

The comparison of MD data calculated with Tersoff~III potential with the results of analytical simulation can be made from the Fig.~\ref{k_vs_Temperature+Analytic}. As one can see, there is qualitative correlations between results of both approaches. For analytical calculations the DOS and dispersion curves was directly taken from the MD simulations (Fig.~\ref{DOS_vs_Strain} and Fig.~\ref{Disp_vs_Strain}, respectively), which were performed also with Tersoff~III potential. The phonon lifetimes were firstly calculated with Eq.~\ref{tau_phonon} for Tersoff~III potential, and then, because of data scattering, they were fitted with Eq.~\ref{fittau}. The calculated lifetimes with the fitting curves are presented on the Fig.~\ref{Phonon_lifetimes}. Fitting curves were used for analytical calculations of thermal conductivity with KT analytical approach.

The quantitative correlations is observed only in the temperature range from 500 to 900 K. Some mismatch in the low temperature range (below Debye temperature) can be explained by classical nature of MD simulations; i.e., the mean number of phonons does not depend on their vibration frequency in this case. Using quantum correction (see Appendix~\ref{quantum_correction}) mostly give general trends. However, thermal conductivity could not be fully corrected with the latter approach~\cite{Turney2009}. Thus, for low temperature range, simulations approaches based on the KT simulations are more reliable. Some mismatches occurring at higher temperatures correspond to the increasing role of higher-order inharmonic phenomena in the strained silicon, because we get close to the melting temperature.

\section{Conclusions}
In this study, we considered thermal transport in crystalline silicon under different magnitude of hydrostatic strains. Particularly, we considered Tersoff~III, Stillinger-Weber (SW), Environment-Dependent Interatomic Potential (EDIP), and Second Nearest Neighbour (2NN) Modified Embedded Atom Method (MEAM) potentials for thermal conductivity evaluation of strained silicon with the use of equilibrium molecular dynamics approach. Firstly, the systems were created by expanding and compression of the crystalline silicon, during this step strain-stress curves were extracted for all potentials. All potentials give the similar value of the bulk modulus (slope of the strain-stress curve). However, only Tersoff potential gives dependence of the thermal conductivity on the strain, which corresponds well with experimental data. Thus, this potential was used further for calculation of dispersion curves, density of states and relaxation time for obtaining physical insight regarding the thermal conductivity behavior of strained silicon.

In addition, these data were used in analytic model based on the kinetic theory. In such way, the dependence of thermal conductivity on the temperature were calculated from analytical relations and correlated with the ones evaluated with equilibrium molecular dynamics approach. Both dependencies  revealed good qualitative agreement. Nevertheless, some quantitative mismatches were observed for low and high temperatures. Mismatch at low temperatures may arise because quantum corrections (see Appendix~\ref{quantum_correction} for more details on the latter correction). At the same time the discrepancies for high temperatures may arise as we get  close to the melting temperature.

This results bring new insights  about strained silicon thermal properties. In future work, the modeling of nanostructures, which exhibits such stained area could be of great interest to tailor phonon transport properties.

\begin{acknowledgments}
This work has been partially funded by the CNRS Energy unit (Cellule Energie) through the project ImHESurNaASA. We want to acknowledge the partial financial support of the scientific pole EMPP of the University of Lorraine. The publication contains the results obtained in the frames of the research work "Features of photothermal and photoacoustic processes in low-dimensional silicon-based semiconductor systems" (Ministry of Education and Science of Ukraine, state registration number 0118U000242).
We wish to acknowledge Dr. Vladimir Lysenko (INL, INSA Lyon) for fruitful discussions.
\end{acknowledgments}

\appendix

\section{Potential models}\label{Potentials}
\subsection{Stillinger-Weber potential}
The SW potential is written in the following form~\cite{Stillinger1985}:
\small
\begin{equation}
	U_{tot}=\sum_{i<j}\phi_{2}(r_{ij})+\sum_{i<j<k}\phi_{3}(r_{ij},r_{ik},\theta_{ijk}),
\end{equation}
\begin{equation}
	\phi_{2}=A_{ij}\epsilon_{ij}\left[B_{ij}\left(\frac{\sigma_{ij}}{r_{ij}}\right)^{p_{ij}}\right]e^{\frac{\sigma_{ij}}{r_{ij}-a_{ij}\sigma_{ij}}},
\end{equation}
\begin{eqnarray}
	\phi_{3}=\lambda_{ijk}\epsilon_{ijk}\left[cos\theta_{ijk}-cos\theta_{0ijk}\right]^{2}\times \nonumber \\
	\times e^{\frac{\gamma_{ij}\sigma_{ij}}{r_{ij}-a_{ij}\sigma_{ij}}}e^{\frac{\gamma_{ik}\sigma_{ik}}{r_{ik}-a_{ik}\sigma_{ik}}},
\end{eqnarray}
\normalsize
The corresponding parameters for the SW potential are provided in table~\ref{SW_Table}.
\begin{table}[h]
	\centering
	\caption{Stillinger-Weber potential parameters for silicon~\cite{Stillinger1985}}\label{SW_Table}
	\begin{tabular}{|c c|c c|}
		\hline
		$\epsilon$ (eV) & 2.1683       & $\sigma$ ({\AA}) & 2.0951            \\
		$a$             & 1.80         & $\lambda$        & 21.0              \\
		$\gamma$        & 1.20         & $\cos\theta_{0}$ & -0.333333333333   \\
		$A$             & 7.049556277  & $B$              & 0.6022245584      \\
		$p$             & 4.0          &                  &                   \\
		\hline
	\end{tabular}
\end{table}

\subsection{Environment-Dependent Interatomic Potential}
The EDIP potential is written as~\cite{Justo1998}
\small
\begin{equation}
	U_{tot}=\sum_{i<j}\phi_{2}(r_{ij},Z_{i})+\sum_{i<j<k}\phi_{3}(r_{ij},r_{ik},Z_{i}),
\end{equation}
\begin{equation}
	\phi_{2}=A\left[\frac{B}{r_{ij}}^{\rho}-e^{-\beta Z_{i}^{2}}\right]e^{\frac{\sigma}{r_{ij}-a}},
\end{equation}
\begin{equation}
	\phi_{3}=e^{\frac{\gamma}{r_{ij}-a}}e^{\frac{\gamma}{r_{ik}-a}}h(\cos\theta_{ijk},Z_{i}),
\end{equation}
\begin{equation}
	Z_{i}=\sum_{m\neq i}f(r_{im}), f(r)= \begin{cases}
		1 & r<c\\
		e^{\frac{\alpha}{1-x^{3}}} & c<r<a \\
		0 & r> a
	\end{cases},
\end{equation}
\begin{equation}
	h(l,Z)=\lambda\left[1-e^{Q(Z)(l+\tau(Z))^{2}}+\eta Q(Z)(l+\tau(Z))^{2}\right],
\end{equation}
\begin{equation}
	Q=Q_{0}e^{-\mu Z}, \tau=u_{1}+u_{2}\left(u_{3}e^{-u_{4}Z}-e^{-2u_{4}Z}\right)
\end{equation}
\normalsize
The corresponding parameters for the EDIP potential are provided in table~\ref{EDIP_Table}.
\begin{table}[h]
	\centering
	\caption{EDIP potential parameters for silicon~\cite{Justo1998}}\label{EDIP_Table}
	\begin{tabular}{|c c|c c|}
		\hline
		$A$ (eV)        & 7.982173        & $B$ ({\AA})         & 1.5075463  \\
		$a$ ({\AA})     & 3.121382        & $c$ ({\AA})         & 2.5609104  \\
		$\alpha$        & 3.1083847       & $\beta$             & 0.0070975  \\
		$\eta$          & 0.2523244       & $\gamma$ ({\AA})    & 1.1247945  \\
		$\lambda$ (eV)  & 1.453310        & $\mu$               & 0.6966326  \\
		$u_{1}$         & 1.1247945       & $\sigma$ ({\AA})    & 0.5774108  \\
		$u_{2}$         & 1.4533108       & $Q_{0}$             & 0.2523244  \\
		$u_{3}$         & 0.6966326       & $\rho$              & 1.2085196  \\
		$u_{4}$         & 1.2085196       &                     &            \\
		\hline
	\end{tabular}
\end{table}

\subsection{Modified Embedded Atom Method Potential}
In the 2NN MEAM formalism~\cite{Lee2000, Lee2001}, the total energy of a system is given by
\small
\begin{equation}
	U_{tot}=\sum_{i}\left[F_{i}(\overline{\rho}_{i})+\frac{1}{2}\sum_{j\neq i}S_{ij}\varphi_{ij}(R_{ij})\right],
\end{equation}
\begin{equation}
	F(\overline{\rho}_{i})=AE_{c}(\overline{\rho}_{i}/\overline{\rho}_{i}^{0})\ln(\overline{\rho}_{i}/\overline{\rho}_{i}^{0}),
\end{equation}
\begin{equation}
	\overline{\rho}_{i}=\overline{\rho}_{i}^{(0)}G(\Gamma),
\end{equation}
\begin{equation}
	G(\Gamma)=\frac{2}{1+e^{-\Gamma}}, \Gamma=\sum_{k=1}^{3}t_{i}^{(k)}\left( \frac{\overline{\rho}_{i}^{(k)}}{\overline{\rho}_{i}^{(0)}} \right)^{2},
\end{equation}
\begin{equation}
	(\overline{\rho}_{i}^{(0)})^{2}=\left[\sum_{j\neq i}\rho_{j}^{a(0)}\right]^{2},
\end{equation}
\begin{equation}
	(\overline{\rho}_{i}^{(1)})^{2}=\sum_{\alpha}\left[\sum_{j\neq i}\frac{r_{ij}^{\alpha}}{r_{ij}}\rho_{j}^{a(1)}\right]^{2},
\end{equation}
\begin{equation}
	(\overline{\rho}_{i}^{(2)})^{2}=\sum_{\alpha,\beta}\left[\sum_{j\neq i}\frac{r_{ij}^{\alpha}r_{ij}^{\beta}}{r_{ij}^{2}}\rho_{j}^{a(2)}\right]^{2}-\frac{1}{3}\left[\sum_{j\neq i}\rho_{j}^{a(2)}\right]^{2},
\end{equation}
\begin{equation}
	(\overline{\rho}_{i}^{(3)})^{2}=\sum_{\alpha,\beta,\gamma}\left[\sum_{j\neq i}\frac{r_{ij}^{\alpha}r_{ij}^{\beta}r_{ij}^{\gamma}}{r_{ij}^{3}}\rho_{j}^{a(3)}\right]^{2}-\frac{3}{5}\sum_{\alpha}\left[\sum_{j\neq i}\frac{r_{ij}^{\alpha}}{r_{ij}}\rho_{j}^{a(3)}\right]^{2},
\end{equation}
\begin{equation}
	\rho_{i}^{a(k)}(r_{ij})=e^{-\beta_{i}^{(k)}(r_{ij}/r_{e}-1)},
\end{equation}
\begin{equation}
	E^{u}(r)=-E_{c}(1+a^{*}+d(a^{*})^{3}),
\end{equation}
\begin{equation}
	a^{*}=\alpha(r/r_{e}-1),
\end{equation}
\begin{equation}
	\alpha=\sqrt{9B\Omega/E_{c}}
\end{equation}
\normalsize
The corresponding parameters for the 2NN MEAM potential are provided in table~\ref{MEAM_Table}.
\begin{table}[h]
	\centering
	\caption{MEAM potential parameters for silicon~\cite{BJLee2007}}\label{MEAM_Table}
	\begin{tabular}{|c c|c c|}
		\hline
		$A$               & 0.58            & $B$ (dyne/cm$^{2}$)   & 0.99$\cdot10^{12}$   \\
		$r_{e}$ ({\AA})   & 2.35            & $E_{c}$ (eV)        & 4.63     \\
		$\beta^{(0)}$     & 3.55            & $t^{(0)}$           & 1.0     \\
		$\beta^{(1)}$     & 2.57            & $t^{(1)}$           & 1.8     \\
		$\beta^{(2)}$     & 0.0             & $t^{(2)}$           & 5.25    \\
		$\beta^{(3)}$     & 7.5             & $t^{(3)}$           & -2.61   \\
		$C_{max}$         & 2.8             & $C_{min}$           & 1.41  \\
		\hline
	\end{tabular}
\end{table}

\subsection{Tersoff Potential}
The Tersoff potential describes the atomic interaction as follows~\cite{Tersoff1989,Tersoff1990}:
\small
\begin{equation}
	U_{tot}=\sum_{i<j}f_{C}(r_{ij})\left[f_{R}(r_{ij})+b_{ij}f_{A}(r_{ij})\right],
\end{equation}
\begin{equation}
	f_{C}(r)=\begin{cases}
		1 & r< R-D \\
		\frac{1}{2}-\frac{1}{2}\sin\left(\frac{\pi(r-R)}{2D}\right) & R-D < r < R+D \\
		0 & r > R+D
	\end{cases},
\end{equation}
\begin{equation}
	f_{A}(r)=-Be^{-\lambda_{2}r},
\end{equation}
\begin{equation}
	b_{ij}=\left(1+\beta^{n}\zeta_{ij}^{n} \right)^{-\frac{1}{2n}},
\end{equation}
\begin{equation}
	\zeta_{ij}=\sum_{k\neq i,j}f_{C}(r_{ik})g(\theta_{ijk})e^{\lambda_{3}^{m}\left(r_{ij}-r_{ik}\right)^{m}},
\end{equation}
\begin{equation}
	g(\theta)=1+\frac{c^{2}}{d^{2}}-\frac{c^{2}}{d^{2}+(\cos\theta-cos\theta_{0})^{2}},
\end{equation}
\normalsize
The corresponding parameters for the Tersoff potential are provided in table~\ref{Tersoff_Table}.
\begin{table}[h]
	\centering
	\caption{Tersoff potential parameters for silicon~\cite{Tersoff1989,Tersoff1990}}\label{Tersoff_Table}
	\begin{tabular}{|c c|c c|}
		\hline
		$m$                              & 3.0                  &  $n$                & 0.78734                      \\
		$R$ ({\AA})                      & 2.85                 &  $D$ ({\AA})        & 0.15                         \\
		$c$                              & 1.0039$\cdot10^{5}$  &  $d$                & 16.217                       \\
		$A$ (eV)                         & 1830.8               &  $B$ (eV)           & 471.18                       \\
		$\lambda_{1}$ ({\AA}$^{-1}$)     & 2.4799               &  $\beta$            & 1.1$\cdot10^{-6}$            \\
		$\lambda_{2}$ ({\AA}$^{-1}$)     & 1.7322               &  $\cos\theta_{0}$   & -0.59825                     \\
		$\lambda_{3}$ ( {\AA}$^{-1}$)    & 0.0                  &                     &                              \\
		\hline
	\end{tabular}
\end{table}

\section{Impact of a cell number in simulation domain on stress-strain curves} \label{strain-stress-dif-cells}
\begin{figure}[h]
	\includegraphics[width=\columnwidth]{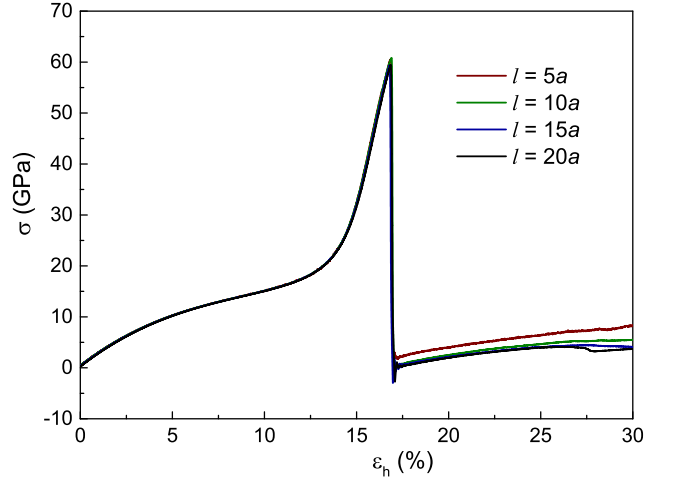}
	\caption{\label{str-stress-diff-cells} Impact of the cell number in simulation domain ($l$) on the strain-stress curves for the Tersoff~III potential.}
\end{figure}

The several sizes of simulation domain ($l$) were checked to find appropriate ones for further simulations. As an example, Fig.~\ref{str-stress-diff-cells} presented the results of simulations of stress-strain curves for Tersoff~III potentials~\cite{Tersoff1988, Tersoff1989, Tersoff1990}. The situation when the simulations domain has number of cells equal to 5, 10, 15, and 20 are demonstrated. In the Fig.~\ref{str-stress-diff-cells}, $a$ is the lattice parameter.

\section{Parameters of fitting of phonon lifetime}\label{fit-phl}
Phonon lifetimes evaluated with MD (Eq.~\ref{tau_phonon}) were fitted with the following equation:
\begin{equation}
	\tau(\omega, p) = A^{-1} \omega^{-\chi} T^{-\xi} \exp({-B/T})
\end{equation}
Parameters of the fitting are presented in the following table.

\begin{table}[h]
	\centering
	\caption{Parameters of fitting for phonon lifetime} \label{Fit_Table}
	\begin{tabular}{|c | c | c | c | c|}
		\hline
		$ $							   & $A (s^{-1-\chi} K^{-\xi})$	&$\xi$ (units) &$\chi$ (units) & $B (K)$\\
		compressed                      & $1.52321\cdot10^{-14}$          &1.60416   & 1.08586   & 0      \\
		unstrained			           & $1.73548\cdot10^{-16}$          &1.76149   & 1.03746   & 0       \\
		tensed                          & $2.87935\cdot10^{-18}$  		&1.91305   & 0.974069  & 0       \\
		\hline
	\end{tabular}
\end{table}

\section{Impact of quantum correction}\label{quantum_correction}
Molecular dynamics operates with classical laws of motion to evaluate the systems with a large number of degrees of freedom. Thus, quantum phenomena are naturally excluded from the consideration in this case. One of the commonly used approach for taken them into account is based on the MD temperature rescaling to the temperature which includes quantum phenomena~\cite{SOLEIMANI2018346}. In this case, thermal conductivity may be corrected as follows~\cite{Lee1991}:

\begin{equation}\label{quantcor}
	\kappa_{qc} = \kappa_{MD}\cdot\frac{dT_{MD}}{dT},
\end{equation}
where $\kappa_{qc}$ is the quantum corrected thermal conductivity, $\kappa_{MD}$ is the thermal conductivity evaluated with MD, $T$ is the temperature, and $T_{MD}$ is the classical temperature, which can be calculated as follows:
\begin{equation}\label{tMD}
	T_{MD} = \frac{T^2}{T_D}\cdot\int_0^{\frac{T_D}{T}}\frac{xdx}{\exp(x)-1} + \frac{T_D}{4},
\end{equation}
where $T_D$ is the Debye temperature.

\begin{figure}[h]
	\includegraphics[width=\columnwidth]{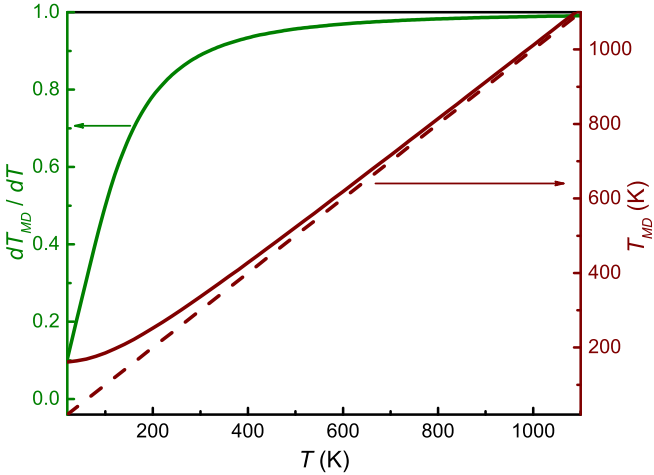}
	\caption{\label{quantcorterm} Dependence of $dT_{MD}/dT$ (left scale) and $T_{MD}$ (right scale) on the corrected temperature. For comparison with the dotted line the dependence $T(T)$ for right scale is presented.}
\end{figure}

The Fig.~\ref{quantcorterm} (left scale) demonstrates temperature dependence of the derivative $\frac{dT_{MD}}{dT}$. This derivative defines the ratio of quantum corrected thermal conductivity to MD one (Eq. \ref{quantcor}). And as one can see from the figure, the quantum corrected thermal conductivity is less than MD one approximately to 10~\% at the 300~K.  Close to Debye temperature ($T_D$ = 658 K for silicon) the difference is less than 3~\%, and only after 1000 K the value of the derivative is close to one. Thus, the impact of the quantum phenomena is significant for the temperatures less than $T_D$.

On the right scale of Fig.~\ref{quantcorterm} we presented dependence of MD temperature on the quantum corrected one. We presented the line which corresponds to the case $T(T)$ with the dotted line.

\begin{figure}[h]
	\includegraphics[width=\columnwidth]{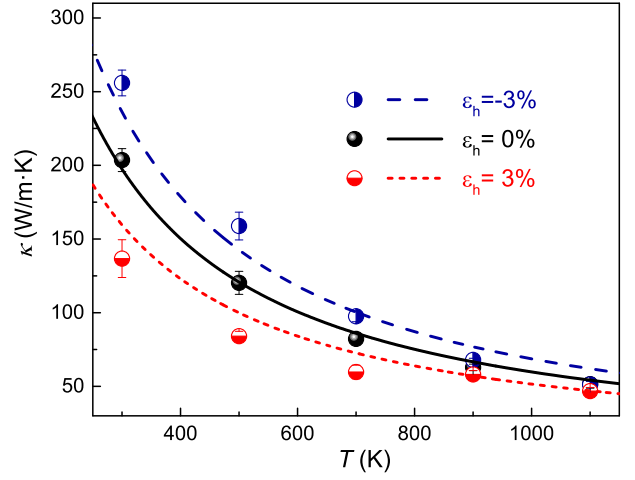}
	\caption{\label{qcTC} Strain-dependent quantum corrected thermal conductivity of silicon between 300 and 1100~K for Tersoff~III potential.}
\end{figure}

The dependence of the quantum corrected in such way thermal conductivity  is presented on the Fig.~\ref{qcTC}. In the correction we neglected change of Debye temperature in strained silicon. As one can see, that we achieve an excellent agreement of MD data with analytical approach. Nevertheless, the analytic model overestimate MD data for compressed silicon and underestimate for tensile one in the low temperature range. Such mismatch can be partially overcome with the correction of Debye temperature, which arise mainly with slightly increasing of sound velocity (see Fig.~\ref{strain_stress}) in compressed silicon and decreasing in tensile one.

Finally, we should note that the use of the quantum correction in the presented above form is only illustrative, and thermal conductivity could not be fully corrected, see for example results of quantum and classical lattice dynamics simulations presented by Turney et al ~\cite{Turney2009}. Therefore, we decided to leave the part with the quantum correction only in the Appendix. Nevertheless, the presented results can be important for understanding of mismatch of kinetic theory analytic approach and MD data without quantum correction.

\bibliography{straineSiTC}

\end{document}